\documentclass[manuscript, nonacm]{acmart}

\usepackage{enumitem}
\AtBeginDocument{%
  }

\setcopyright{acmlicensed}
\copyrightyear{2026}
\acmYear{2026}
\acmDOI{XXXXXXX.XXXXXXX}
\acmISBN{978-1-4503-XXXX-X/2026/09}

\setcopyright{none}

\begin{document}

\title{Detection-Efficient Watermarking for Vision--Language Agents via Risk-Constrained Strength Allocation}


\author{Dumindu Samaraweera}
\correspondingauthor
\email{samarawg@erau.edu}
\orcid{0000-0003-4097-5585}

\author{Prashant Shekhar}
\email{shekharp@erau.edu}
\orcid{0000-0003-2353-6740}
\affiliation{%
  \institution{Embry-Riddle Aeronautical University}
  \city{Daytona Beach}
  \state{Florida}
  \country{USA}
}

\renewcommand{\shortauthors}{Samaraweera et al.}

\begin{abstract}
Vision--language agents generate descriptions, decisions, and instructions that
may be reused by users, downstream agents, and automated workflows, creating a
need for provenance mechanisms that support content authenticity and
post-deployment auditing without unnecessarily degrading response quality. We
present a risk-aware semantic watermarking framework that combines predictive
uncertainty with a zero-shot multimodal relevance-risk proxy to allocate
watermark strength during decoding. A single risk penalty $\lambda$ exposes a
continuous operating frontier, allowing deployment-time selection of a
detection--quality trade-off without model retraining. At a selected operating
point, mean watermark strength is reduced by approximately 59.4\% relative to
fixed-strength watermarking, while perplexity change falls from 0.3634 to
0.1689 and clean/paraphrased ROC-AUC reaches 0.9552/0.9459. A matched-strength
analysis shows that this advantage reflects the allocation policy rather than
lower average intervention: at comparable mean strength, risk-constrained
allocation achieves substantially higher ROC-AUC than fixed, entropy-only, and
criticality-only policies, identifying it as detection-efficient rather than
quality-dominant. We further show that threshold calibration on finite samples
produces substantial false-positive-rate drift, motivating calibration
variance reporting as standard practice for auditable provenance. 
\end{abstract}




\maketitle

\section{Introduction}
Vision--language agents are increasingly used to interpret images, answer visual questions, generate descriptions, and support decisions in interactive systems. Their outputs may be copied into reports, consumed by downstream agents, or treated as evidence in operational workflows. These uses create a need for mechanisms that help determine whether an output originated from an authorized model and whether it has been modified after generation.

Statistical watermarking provides a widely used provenance mechanism. During decoding, a watermarking procedure modifies the token distribution according to a secret rule, and a detector then evaluates whether the resulting response contains statistical evidence of the watermark. This supports provenance verification, content authenticity, auditing, and incident investigation without requiring model retraining. However, watermarking introduces a security--quality trade-off: stronger perturbations generally improve detection, but they can also alter word choice, likelihood, fluency, and task-relevant content. Existing schemes apply a fixed watermark strength to every token decision, which is reasonable for open-ended text where a changed token is usually one plausible continuation among many. It is a poor fit for vision--language agents, whose responses may contain visually grounded objects, attributes, spatial relations, and action recommendations: modifying a token that expresses a critical visual relation may be far more harmful than modifying a stylistic or redundant one, yet a uniform-strength policy treats both decisions identically.

A small number of recent methods have begun to address this limitation by adapting watermark strength at decoding time. Entropy-based schemes increase watermark pressure when the model's predictive distribution is dispersed and reduce it when the distribution is concentrated, which allows the watermark to exploit positions where the model has more freedom. Vision--language watermarking extends this idea to multimodal generation. These approaches are promising, but they share two limitations. First, they base allocation on generation uncertainty alone, without any signal about whether the current decision is semantically important to the response or grounded in the input image. Second, they are typically evaluated at a single default configuration, so it is unclear whether reported gains reflect the allocation policy or simply a lower average intervention.

This paper studies risk-aware allocation of watermark strength during vision--language-agent decoding. At each step, the proposed method combines two signals: predictive uncertainty, represented by normalized token-distribution entropy, and a zero-shot multimodal relevance-risk proxy derived from image-conditioned and language-context representations. Watermark strength is increased when the model has greater freedom among plausible tokens, and reduced when the current probability mass is concentrated on tokens the proxy considers relevant to the visual or linguistic context. The resulting policy is parameterized by a single risk penalty \(\lambda\), which controls a continuous family of allocation behaviors from near-fixed watermarking to strongly attenuated watermarking while holding the model, partition mechanism, and detector fixed. The proxy is deliberately treated as an operational risk signal rather than a calibrated estimate of semantic importance; the paper evaluates it empirically rather than assuming it is correct.

Our evaluation is designed to separate two questions that are often conflated in adaptive-watermarking work: whether a method improves the detection--quality trade-off at its default configuration, and whether it improves that trade-off through the allocation policy rather than through lower average intervention. To address the second question we introduce a matched-strength analysis that compares methods at comparable mean watermark strength, isolating the effect of where the watermark budget is spent from the effect of how much of it is spent. We also examine deployment realism: because a detector threshold must be calibrated on finite unwatermarked samples, we characterize how much the empirical false-positive rate drifts across calibration splits, and we argue that calibration variance should be reported as standard practice for auditable provenance systems. In summary, this paper makes three contributions:

\begin{itemize}
    \item We formulate adaptive VLA watermarking as a \emph{detection--quality allocation problem} parameterized by a single risk penalty \(\lambda\), give an explicit implementation-level definition of the multimodal relevance-risk proxy, and derive the strength-allocation rule from a local detection-benefit versus risk objective.
    \item We evaluate fixed-strength, entropy-only, criticality-only, and joint risk-constrained strategies under a common model, detector, prompt configuration, and 1,000-instance AMBER evaluation set, and we introduce a \emph{matched-strength analysis} that separates gains attributable to the allocation policy from gains attributable to lower average intervention.
    \item We characterize the calibration behavior of the detector across repeated held-out splits and argue that finite-sample false-positive drift should be reported as standard practice for auditable provenance systems.
\end{itemize}

The scope of our claims is deliberately narrow. We do not claim that risk-aware allocation dominates fixed or entropy-only watermarking on every metric. We do not claim that the zero-shot relevance proxy is a validated semantic-importance estimator. We do not claim robustness beyond the single synonym-substitution and paraphrase procedures we tested. What we do claim is that risk-aware allocation exposes a \emph{controllable and detection-efficient operating frontier} for VLA provenance, that this frontier is not reachable by fixed or entropy-only policies at equal watermark budget, and that deployment requires explicit calibration management. These claims are consistent with, and complementary to, the direction explored by VLA-Mark \cite{liu2025vla}; a same-environment numerical comparison is left to future work.

The remainder of the paper is organized as follows. Section~2 motivates the security--trustworthiness tension. Section~3 reviews related work. Section~4 defines the threat model and design goals. Section~5 presents the risk-aware watermarking method and derives the allocation rule. Section~6 describes the experimental setup. Section~7 reports detection, quality, matched-strength, and calibration results. Sections~8 and~9 discuss implications, limitations, and responsible-use considerations.

\section{Security and Trustworthiness Motivation}
The security objective of watermarking is to make generated content statistically distinguishable from unwatermarked content. A verifier with access to the detector and secret partitioning procedure should be able to estimate whether a response was produced by the marked generation process. However, the trustworthiness objective is broader. A provenance mechanism should avoid unnecessarily changing factual content, visual grounding, fluency, and user utility. These objectives can conflict because stronger watermarking usually increases the statistical signal while also increasing the likelihood of changing the generated response.

In vision-language settings, for a visual description, some tokens may express important objects, attributes, or spatial relations, while others may be stylistic or redundant. A uniform watermark strength treats these tokens identically. A risk-aware policy instead attempts to place stronger watermark pressure where the expected semantic effect is lower.

This perspective treats watermarking as part of an auditable deployment workflow. Detection metrics characterize provenance performance, while quality metrics characterize the potential cost of applying the provenance mechanism. It is important to note that, neither component is sufficient in isolation. A detector with excellent AUC but severe output degradation may be unsuitable for deployment, whereas a nearly invisible watermark with weak detection may provide little practical accountability. A practical consequence of this tension is that a useful provenance mechanism should expose an \emph{operating frontier} rather than a single fixed configuration, so that deployment teams can choose a detection--quality point appropriate to their auditing requirements. The method developed in Section~\ref{sec:method} is designed around this requirement.

\section{Related Work}
\subsection{Statistical Text Watermarking}
Many text watermarking systems partition the vocabulary into green and red lists using a secret key and modify the decoding distribution to favor green-list tokens~\cite{kirchenbauer2023watermark}. Detection then measures whether the response contains more green-list tokens than expected under unwatermarked generation. Alternative distortion-free and distribution-preserving schemes avoid explicitly biasing the token distribution~\cite{aaronson2024watermarking, kuditipudi2023robust, christ2024undetectable}, but share the same basic property: the watermark is inserted into the generated token sequence and recovered by a detector that has access to the secret rule.

These methods are attractive because they do not require model retraining and can be integrated directly into decoding. Their limitation is that a fixed watermark bias applies the same intervention to tokens with different semantic importance. Adaptive methods address this limitation by varying the partition, bias, or strength according to context, token probability, or model uncertainty~\cite{kirchenbauer2023watermark}. This work belongs to this adaptive family and studies how allocation strength should depend on multimodal context in a vision--language agent.

\subsection{Watermarking for Vision--Language Models}
In a vision--language model, the image affects the language-model logits through visual conditioning. The watermark is still inserted into the generated token sequence, but the visual context provides additional information that may be used when allocating watermark strength~\cite{liu2025vla}. This creates an opportunity to distinguish between tokens that are likely to express image-grounded content and tokens that are less central to the response. It also introduces additional evaluation requirements. Textual fluency alone does not establish that the response remains consistent with the input image, and a watermark that changes visually grounded content may be more harmful than one that changes stylistic tokens.

\subsection{VLA Watermarking and the VLA-Mark Baseline}
VLA-Mark is the closest prior comparison to this work because it applies watermarking to vision--language generation and uses entropy-related information to regulate the watermarking process~\cite{liu2025vla}. Its allocation is driven by generation uncertainty: watermark pressure is adjusted according to how concentrated or dispersed the model's predictive distribution is. It does not incorporate a token-level multimodal relevance signal that attenuates intervention when probability mass concentrates on visually or linguistically grounded content. The present work extends this direction by introducing such a signal and by explicitly studying matched-strength operating points that separate gains attributable to the allocation policy from gains attributable to lower average intervention.

Published VLA-Mark results are not treated as directly comparable numerical baselines, because differences in model checkpoints, prompt templates, decoding settings, datasets, detector thresholds, and evaluation splits can shift reported metrics substantially. Comparing against published numbers without a common environment would conflate allocation-policy effects with implementation effects, which is precisely the kind of comparison that undermines reproducible evaluation. This paper therefore focuses on controlled within-environment comparisons; a same-environment reproduction of VLA-Mark is a necessary next step, and we make our implementation and evaluation protocol explicit so that such a comparison can be run directly.

Accordingly, this paper focuses on controlled comparisons within a common implementation. The main question is not whether the proposed method dominates every prior system, but whether its allocation rule exposes a useful and reproducible security--quality frontier (Section~\ref{sec:results}), and whether that frontier is reachable by simpler allocation policies at equal watermark budget (Section~\ref{sec:matched-strength}). These questions can be answered without a cross-implementation comparison, and the answers constrain how risk-aware allocation should be used in trustworthy deployment.

\section{Threat Model and Design Goals}
We consider a model provider that controls a vision--language generation process and inserts a secret watermark into generated responses during decoding, conditioned on both the input image and the user query. The provider also controls the corresponding detector and key-dependent vocabulary partitioning. A verifier uses the detector to assess whether a response is likely to have originated from the marked model.

The primary threat is unauthorized use or misrepresentation of generated content. The watermark is intended to provide statistical provenance evidence after generation. It is not intended to prove factual correctness, safety, authorization, or privacy preservation. Transformations such as paraphrasing, synonym substitution, deletion, truncation, and adversarial rewriting may weaken token-level watermark evidence. We therefore treat transformation robustness as a separate evaluation dimension. The current experiments evaluate one synonym-substitution procedure and one paraphrase procedure. Thus, for this work, the design goals are:

\begin{enumerate}
    \item retain a statistically detectable watermark;
    \item reduce unnecessary perturbation of semantically sensitive decisions;
    \item preserve linguistic and image--text quality;
    \item expose an interpretable security--quality operating frontier whose points can be selected at deployment time; and
    \item support threshold calibration suitable for post-deployment auditing, while reporting calibration variance rather than a single nominal operating point.
\end{enumerate}

\section{Risk-Aware Semantic Watermarking}\label{sec:method}
The proposed methodology is designed around the operating-frontier requirement of Section~2: a single scalar policy parameter $\lambda$ should traverse a family of allocation policies, from near-fixed behavior at small $\lambda$ to strongly attenuated watermarking at large $\lambda$, without changing the model, the detector, or the key.

Let $V$ be the model vocabulary and let $z_t(v)$ denote the logit for candidate token $v$ at decoding step $t$. The unwatermarked token distribution is
\begin{equation}
p_t(v)=\operatorname{softmax}(z_t(v)).
\end{equation}

A secret pseudo-random procedure partitions the vocabulary into green and red sets. The partition may depend on the decoding context and a secret key. Green-list tokens contribute positive evidence to the detector, while red-list tokens do not receive the same preference.

\subsection{Uncertainty Signal}
The predictive entropy at decoding step $t$ is
\begin{equation}
H_t=-\sum_{v\in V}p_t(v)\log p_t(v).
\end{equation}

We normalize entropy to obtain
\begin{equation}
g_t=\operatorname{Norm}(H_t), \qquad g_t\in[0,1].
\end{equation}

Large $g_t$ indicates that probability mass is distributed over multiple plausible tokens. Such a distribution provides greater freedom to bias generation toward the green list. Low $g_t$ indicates a more concentrated distribution, where modifying the preferred token may have a larger effect on the response.

\subsection{Multimodal Relevance-Risk Proxy}
Let $e_v$ denote the embedding of vocabulary token $v$. Let $H_L=\{h_i^L\}$ denote language-context representations and let $h_{\mathrm{img}}$ denote an image-conditioned representation. We define context, local-variation, and image relevance scores as
\begin{align}
s^{\mathrm{ctx}}(v)
&=
\operatorname{softmax}_v
\left(
\max_i \operatorname{sim}(h_i^L,e_v)
\right),\\
s^{\mathrm{local}}(v)
&=
\operatorname{softmax}_v
\left(
\max_i \operatorname{sim}(h_i^L,e_v)
-
\min_i \operatorname{sim}(h_i^L,e_v)
\right),\\
s^{\mathrm{img}}(v)
&=
\operatorname{softmax}_v
\left(
\operatorname{sim}(h_{\mathrm{img}},e_v)
\right).
\end{align}

Intuitively, $s^{\mathrm{ctx}}$ captures how much a candidate token is supported by the language context, $s^{\mathrm{local}}$ captures tokens whose similarity to the language context varies strongly across positions (a proxy for local salience), and $s^{\mathrm{img}}$ captures how much a token is supported by the image representation. The local-variation score is included because tokens with high variance in context similarity tend to be content-bearing rather than purely functional.

In our implementation, $H_L$ is taken as the set of final-layer hidden states of the generated prefix (including the prompt), and $h_{\mathrm{img}}$ is the image feature produced by the vision encoder after projection into the language-model embedding space. Token embeddings $e_v$ are drawn from the same embedding matrix used by the language model. The similarity function $\operatorname{sim}(\cdot,\cdot)$ is cosine similarity.

The vocabulary-level relevance score is
\begin{equation}
r(v)=
\operatorname{Norm}_{\mathrm{mm}}
\left(
s^{\mathrm{ctx}}(v)
+
s^{\mathrm{local}}(v)
+
s^{\mathrm{img}}(v)
\right),
\qquad r(v)\in[0,1].
\end{equation}

Here $\operatorname{Norm}_{\mathrm{mm}}$ denotes min-max normalization over the vocabulary, mapping the summed score into $[0,1]$; $\operatorname{Norm}(\cdot)$ in Eq.~(3) denotes the same min-max normalization applied to the current entropy against a running estimate of the entropy range.

The decision-level risk proxy is the expected relevance under the current model distribution:
\begin{equation}
c_t=\sum_{v\in V}p_t(v)r(v),
\qquad c_t\in[0,1].
\label{eq:risk-proxy}
\end{equation}

A large $c_t$ means that the current probability mass is concentrated on tokens considered relevant by the proxy. It does not imply that a token is certainly necessary or that perturbing it will definitely cause semantic harm.

It is noteworthy that we use the term \emph{relevance-risk proxy} rather than semantic criticality because the current score is deterministic and zero-shot. It is not trained against human judgments, counterfactual task loss, or downstream semantic error.

\subsection{Risk-Constrained Strength Allocation}
Let $\Delta_t$ denote watermark strength, let $\Delta_{\max}$ be its maximum value, and let $\lambda$ control the penalty assigned to relevance risk. We motivate the allocation rule using
\begin{equation}
\max_{0\leq\Delta_t\leq\Delta_{\max}}
\quad
g_t\Delta_t-\lambda(c_t+\epsilon)\Delta_t^2.
\label{eq:allocation-objective}
\end{equation}

The first term rewards watermark strength when detection opportunity is high. The second term penalizes intervention when the relevance-risk proxy is high. The unconstrained stationary point is
\begin{equation}
\Delta_t^\star=
\frac{g_t}{2\lambda(c_t+\epsilon)}.
\end{equation}

Applying the feasible-set constraint gives
\begin{equation}
\Delta_t=
\min\left(
\Delta_{\max},
\frac{g_t}{2\lambda(c_t+\epsilon)}
\right).
\label{eq:adaptive-strength}
\end{equation}

The parameter $\lambda$ is a policy parameter that controls the operating point. However, it is not learned from semantic annotations and should not be interpreted as a calibrated risk coefficient.

The comparison strategies are
\begin{align}
\Delta_t^{\mathrm{fixed}} &= \Delta,\\
\Delta_t^{\mathrm{entropy}} &=
\Delta_{\min}
+
(\Delta_{\max}-\Delta_{\min})g_t,\\
\Delta_t^{\mathrm{criticality}} &=
f(c_t),\\
\Delta_t^{\mathrm{risk}} &=
\min\left(
\Delta_{\max},
\frac{g_t}{2\lambda(c_t+\epsilon)}
\right).
\end{align}

These variants isolate the effects of fixed strength, entropy, relevance risk, and their joint use.

\subsection{Watermarked Decoding and Detection}
At each decoding step, the watermark processor constructs a key-dependent green/red partition and modifies candidate-token scores according to $\Delta_t$. The next token is then sampled or selected using the modified distribution.

The detector computes a score based on the observed green-list membership of generated tokens. A response is classified as watermarked when its score exceeds a selected threshold. We report ROC-AUC to evaluate ranking performance and accuracy, precision, recall, and F1 at selected thresholds to characterize operational behavior.

In our implementation the detector score follows the standard green-list $z$-score: if $G$ is the number of green-list tokens in a response of length $T$ and $\gamma$ is the green-list fraction, the score is $(G - \gamma T)/\sqrt{T\gamma(1-\gamma)}$. A response is classified as watermarked when this score exceeds a threshold $\tau$. Detection metrics in Section~\ref{sec:results} are computed over thresholds $\tau$; ROC-AUC integrates over all $\tau$, while accuracy, precision, recall, and F1 are reported at selected $\tau$ values.

\section{Experimental Methodology}
\subsection{Dataset and Model}
We evaluate all methods on 1,000 image--query instances drawn from the AMBER benchmark \cite{wang2023amber}. Each instance uses the same image and query to generate an
unwatermarked response and a response under each watermarking configuration,
so that comparisons are paired at the instance level. All methods share the
same model, processor, prompt construction, decoding configuration, watermark
keying procedure, detector, and evaluation instances; the experimental
comparison therefore isolates the effect of the strength-allocation policy
alone.

The primary risk-constrained configuration uses $\Delta_{\max}=2.0$ and
$\lambda=0.20$. This operating point was selected on a separate calibration
sweep and was not optimized against the final 1,000-instance test results, so
the reported numbers are not tuned to the evaluation set.

\paragraph{Implementation details.}
We use the public LLaVA-1.5-7B checkpoint~\cite{liu2023llava}
(\texttt{liuhaotian/llava-v1.5-7b}), which pairs a Vicuna-7B-v1.5 language
backbone with a CLIP ViT-L/14-336 vision encoder. Images are loaded as RGB and
processed by the checkpoint's \texttt{AutoProcessor}, which applies the
standard LLaVA image preprocessing for the CLIP ViT-L/14-336 encoder. The
prompt follows the standard LLaVA chat format:
\texttt{USER: <image> \textbackslash n\{question\} ASSISTANT:}. Generation uses
greedy decoding (\texttt{do\_sample=False}, \texttt{num\_beams=1}) with
\texttt{min\_new\_tokens=1} and \texttt{max\_new\_tokens=128}; no temperature
or top-$p$ sampling is used.

Because generation is greedy, it is deterministic given the inputs and model.
The implementation does not set a global random seed: the bootstrap confidence
intervals use seed~$42$, and the synonym-substitution procedure seeds each
instance with its image index. The experiments are therefore best described as
deterministic decoding with controlled seeds for the stochastic evaluation
procedures, rather than as sharing a single global seed across all operations.

The AMBER evaluation data are loaded in JSON order, and the first $1{,}000$
records are used. This is a fixed 1,000-instance subset of the available AMBER
\texttt{query\_generative} records; we refer to it as our evaluation set rather
than as the full AMBER benchmark.

\paragraph{Watermark injection.}
At each decoding step, the watermark processor constructs a key-dependent
green/red vocabulary partition. In our implementation, the partition index is
generated from a deterministic pseudo-random permutation initialized with a
fixed secret key and the most recently generated token, so the effective
prefix length is one token. The green list is populated in two stages: it
first contains a set of semantically ranked tokens whose size depends on the
current normalized entropy $g_t$, and it is then filled with pseudo-randomly
selected vocabulary tokens until it contains approximately half of the
vocabulary. All remaining candidate tokens form the red list.

At decoding step $t$, the adaptive strength $\Delta_t$ is applied by
subtracting $\Delta_t$ from the logits of red-list tokens. Because adding the
same constant to all logits leaves the softmax distribution unchanged, this is
equivalent to adding $\Delta_t$ to green-list logits. The resulting logits are
passed to the same greedy decoder described above. The fixed baseline uses the
same partition construction and logit-bias mechanism but applies a constant
strength $\Delta_t=\Delta$. Within our implementation, the only difference
between the fixed and adaptive baselines is therefore the strength-allocation
policy, which is the effect isolated in Section~\ref{sec:results}.
\subsection{Compared Methods}
We compare:

\begin{enumerate}
    \item fixed-strength watermarking;
    \item entropy-only adaptive watermarking;
    \item criticality-only adaptive watermarking; and
    \item joint risk-constrained watermarking.
\end{enumerate}

In addition to the default comparison, we perform a matched-strength sweep. Fixed, entropy-only, and criticality-only methods are evaluated over several nominal strength values, while the risk-constrained method is evaluated over multiple $\lambda$ values. This analysis compares methods at similar observed mean watermark strength rather than comparing only their default configurations. For fixed, entropy-only, and criticality-only watermarking, we sweep nominal
strength values
\[
\Delta \in \{0.5,\;0.8,\;1.2,\;1.6,\;2.0\}.
\]
For the risk-constrained method, we sweep
\[
\lambda \in \{0.05,\;0.10,\;0.15,\;0.20,\;0.30,\;0.40\},
\]
while keeping $\Delta_{\max}=2.0$. Matched-strength points are obtained by
evaluating each policy at these operating points, recording the observed mean
token-level $\Delta_t$, and comparing policies at the nearest observed mean
strength. The curves in Figure~\ref{fig:matched-budget-frontier} connect these
observed operating points.

\subsection{Metrics}
Detection quality is measured with ROC-AUC, accuracy, precision, recall, and
F1. To evaluate operational behavior at a fixed decision point, accuracy,
precision, recall, and F1 are reported at selected detector thresholds, while
ROC-AUC summarizes ranking quality across all thresholds. The magnitude of the
watermark intervention is summarized by the mean token-level strength
$\Delta_t$.

We evaluate response quality along three complementary dimensions:
response perplexity change relative to the unwatermarked response, response
similarity to the unwatermarked output, and CLIP-based image--text similarity
change. Lower perplexity change indicates less likelihood disruption, and
higher response similarity indicates closer agreement with the unwatermarked
response.

Perplexity is computed by the same LLaVA-1.5-7B model conditioned on the
original image and prompt. The generated response is concatenated with the
original prompt, and the loss is computed only over the generated
response-token positions; prompt positions, including the image-conditioning
portion, are masked with label value $-100$. The model is evaluated without the
watermark logits processor, so the perplexity measurement reflects the
response itself rather than the watermarking intervention.

CLIP-based \cite{radford2021clip} image--text similarity is computed between the input image and the
generated response. Text is chunked into segments of at most 77 tokens, and the
per-chunk CLIP scores are averaged. CLIP change measures only the change in the
selected image--text embedding metric; a near-zero CLIP change is not evidence
of preserved factual correctness or fine-grained visual grounding.

Throughout this paper, perplexity, response similarity, and CLIP are treated as
automatic proxy metrics. They are informative about likelihood, textual
agreement, and image--text alignment under the chosen scorers, but they do not
substitute for human judgments, grounded question-answering accuracy, or
task-specific semantic evaluation.

\subsection{Transformation Robustness}
We evaluate robustness using synonym substitution and the evaluated paraphrase procedure. The detector is applied to transformed watermarked responses using the same detector configuration. These experiments measure robustness to the tested transformations only and do not establish resistance to arbitrary rewriting or adaptive attacks.

Paraphrases are generated using the
\texttt{ramsrigouthamg/t5\_paraphraser} checkpoint through a Hugging Face
\texttt{text2text-generation} pipeline. The input follows the template
\texttt{paraphrase: \{watermarked response\}}. Generation is deterministic,
using \texttt{do\_sample=False} and \texttt{num\_return\_sequences=1}, with a
maximum output length of 128 tokens under the default configuration.
Synonym substitution uses the NLTK WordNet lexical database, with up to five
alphabetic words replaced per response.

Synonym substitution uses the NLTK WordNet lexical database. Up to five alphabetic words are replaced per response using randomly selected WordNet synonyms, excluding candidates identical to the original word. The substitution procedure uses a deterministic per-instance seed based on the image index.

\subsection{Threshold Calibration}
For exploratory calibration, 200 unwatermarked instances are used to select a threshold and 800 instances are held out for evaluation. The calibration targets nominal false-positive rates of $1\%$, $5\%$, and $10\%$. We repeat the 200/800 split $R=20$ times with different random seeds and report mean and standard deviation of the empirical FPR across repetitions.

Because the threshold is estimated from a finite sample, the empirical test FPR may differ from the nominal target. We therefore report both quantities and interpret the calibration results as operating-point estimates rather than deployment guarantees.

\section{Results}
\label{sec:results}

\subsection{Default Detection and Watermark Strength}
\label{sec:default-detection}

Table~\ref{tab:detection} reports detection and mean intervention for all
methods at their default configurations on the full 1,000-instance evaluation
set. The results trace a clear frontier rather than a single ranking. Fixed and
criticality-only watermarking occupy the high-detection, high-intervention end,
with ROC-AUC values of $0.9921$ and $0.9879$ at mean strengths of $2.000$ and
$1.860$. Entropy-only and risk-constrained watermarking occupy the
lower-intervention end, with ROC-AUC values of $0.9065$ and $0.9552$ at mean
strengths of $0.902$ and $0.812$. The risk-constrained configuration achieves
ROC-AUC $0.9552$ at $59.4\%$ lower mean intervention than fixed watermarking,
positioning it on the detection-efficiency side of the frontier.

\begin{table}[t]
    \caption{Default detection and watermark-strength results on 1,000 image--query instances.}
    \label{tab:detection}
    \centering
    \Large
    \setlength{\tabcolsep}{3pt}
    \begin{tabular}{@{}lrrrrr@{}}
    \toprule
    Method & AUC & Accuracy & F1 & Recall & Mean $\Delta$\\
    \midrule
    Fixed & 0.9921 & 0.9335 & 0.9370 & 0.9890 & 2.000\\
    Entropy-only & 0.9065 & 0.8210 & 0.8102 & 0.7640 & 0.902\\
    Criticality-only & 0.9879 & 0.9305 & 0.9340 & 0.9830 & 1.860\\
    Risk-constrained & 0.9552 & 0.8915 & 0.8929 & 0.9050 & 0.812\\
    \bottomrule
    \end{tabular}%
\end{table}

\begin{figure}[t]
    \centering
    \includegraphics[width=0.8\linewidth]{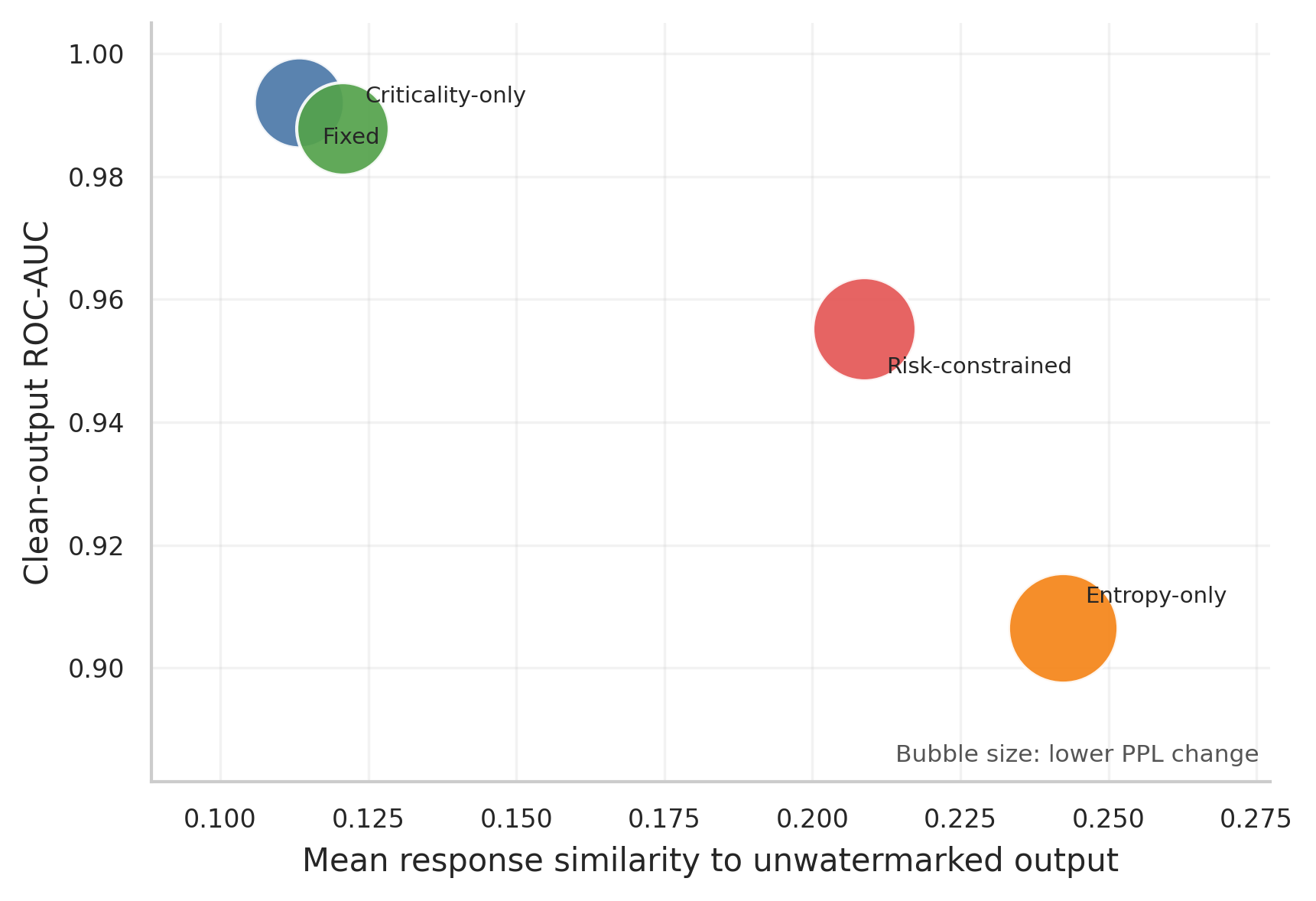}
    \caption{Quality--detection operating points for the default watermarking configurations. The horizontal axis shows response similarity to the unwatermarked output, the vertical axis shows clean-output ROC-AUC, and bubble size represents lower perplexity change.}
    \label{fig:quality-detection-pareto}
\end{figure}

Figure~\ref{fig:quality-detection-pareto} visualizes the default
quality--detection operating points. Fixed and criticality-only watermarking
provide the strongest detection at high intervention; entropy-only and
risk-constrained watermarking trade detection for reduced perturbation. The
risk-constrained point lies between these extremes, consistent with its role as
an intermediate operating point on the frontier.

\subsection{Quality and Image--Text Consistency}
\label{sec:quality}

Table~\ref{tab:quality} reports automatic quality measurements. All changes are
computed relative to the unwatermarked response for the same instance.
Risk-constrained watermarking reduces the mean perplexity change from $0.3634$
(fixed) to $0.1689$ and increases response similarity from $0.1134$ to
$0.2087$. Entropy-only watermarking achieves the strongest automatic quality
values (perplexity change $0.0994$, response similarity $0.2423$), but its
detection ROC-AUC is only $0.9065$—an illustration of the detection-quality
tension the frontier encodes.

\begin{table}[t]
    \caption{Automatic quality results on 1,000 image--query instances. Lower perplexity change is better, while higher response similarity is better.}
    \label{tab:quality}
    \centering
    \Large
    \setlength{\tabcolsep}{2.5pt}
    \begin{tabular}{@{}lrrrrr@{}}
    \toprule
    Method & $\Delta$PPL & $\Delta$CLIP & Similarity & Mean $\Delta$ & Synonym drop\\
    \midrule
    Fixed & 0.3634 & $-0.0042$ & 0.1134 & 2.000 & 0.1758\\
    Entropy-only & 0.0994 & $<0.0001$ & 0.2423 & 0.902 & 0.1939\\
    Criticality-only & 0.3217 & $-0.0029$ & 0.1207 & 1.860 & 0.1722\\
    Risk-constrained & 0.1689 & $+0.0009$ & 0.2087 & 0.812 & 0.2400\\
    \bottomrule
    \end{tabular}%
\end{table}

The CLIP changes are small for all methods. Fixed watermarking changes the
image--text score by $-0.0042$; entropy-only and risk-constrained watermarking
change it by less than $10^{-3}$ in magnitude. These results indicate no
substantial average change under the selected CLIP metric. They should not be
interpreted as proof that visual facts, spatial relations, or downstream task
performance are preserved; they characterize the stability of one automatic
image--text scorer under the watermark intervention.

\begin{figure}[t]
    \centering
    \includegraphics[width=0.8\linewidth]{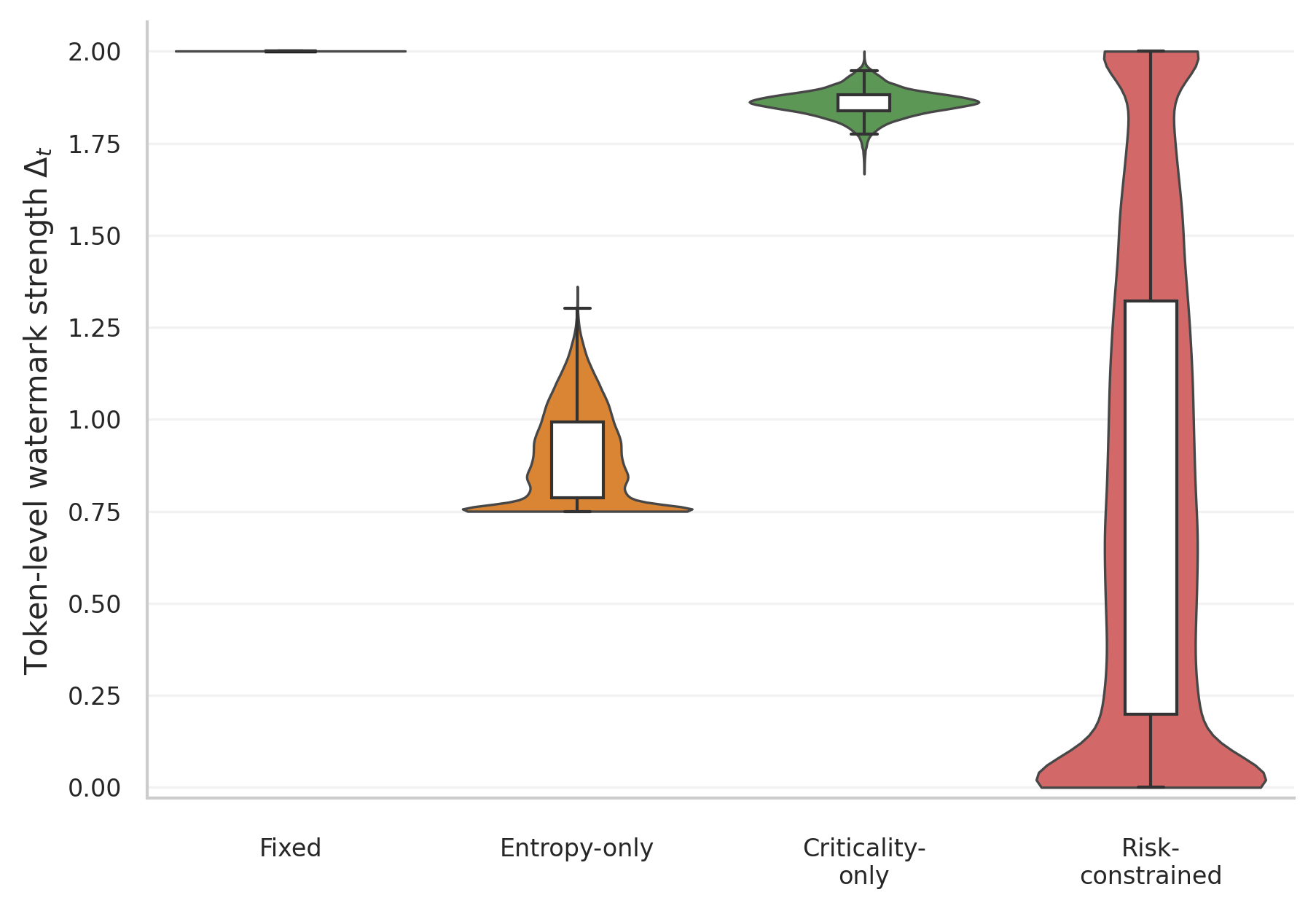}
    \caption{Token-level watermark-strength distributions across the full evaluation set. Adaptive methods redistribute watermark strength across decoding steps rather than applying a uniform intervention.}
    \label{fig:token-strength}
\end{figure}

Figure~\ref{fig:token-strength} shows how adaptive methods redistribute
watermark strength across decoding steps. Fixed watermarking produces a
single-point distribution at $\Delta=2.0$; adaptive policies produce
distributions whose mass shifts toward lower strengths, with the
risk-constrained distribution shifting furthest.

\subsection{Matched-Strength Comparison}
\label{sec:matched-strength}

The default comparison evaluates methods at different observed mean strengths.
We therefore perform a matched-strength sweep using 1,000 instances per
configuration. Fixed, entropy-only, and criticality-only methods are evaluated
over several nominal strength values, and the risk-constrained method is
evaluated over multiple values of $\lambda$. Matched points are obtained by
evaluating each policy at several strengths, recording the observed mean
token-level $\Delta_t$, and comparing policies at the nearest observed mean
strength. For the comparison reported here, the nearest points to mean strength
$0.8$ are $0.800$, $0.756$, $0.744$, and $0.812$ for fixed, entropy-only,
criticality-only, and risk-constrained watermarking, respectively.

Table~\ref{tab:matched-strength} reports the results. At matched mean strength,
risk-constrained allocation achieves ROC-AUC $0.9552$, substantially above
fixed ($0.8553$), entropy-only ($0.8442$), and criticality-only ($0.8405$).
This is the mechanism-level finding: at equal watermark budget, the risk-aware
rule concentrates intervention on token positions where detection evidence is
most recoverable, rather than spreading it uniformly. The same concentration
produces larger automatic quality perturbation (perplexity change $0.1689$ vs.
$0.0600$--$0.0682$), which is the expected cost of reallocating rather than
reducing intervention. Fixed, entropy-only, and criticality-only policies
cannot reach the same detection levels at equal mean strength, because they
have no mechanism for redirecting budget across decisions.

\begin{table}[t]
    \caption{Matched-strength comparison near observed mean watermark strength $0.8$.}
    \label{tab:matched-strength}
    \centering
    \Large
    \setlength{\tabcolsep}{2.5pt}
    \begin{tabular}{@{}lrrrrr@{}}
    \toprule
    Method & Mean $\Delta$ & AUC & $\Delta$PPL & Similarity & $\Delta$CLIP\\
    \midrule
    Fixed ($0.8$) & 0.800 & 0.8553 & 0.0682 & 0.2893 & $+0.0002$\\
    Entropy-only & 0.756 & 0.8442 & 0.0624 & 0.2978 & $+0.0001$\\
    Criticality-only & 0.744 & 0.8405 & 0.0600 & 0.3024 & $+0.0008$\\
    Risk ($\lambda=0.20$) & 0.812 & 0.9552 & 0.1689 & 0.2087 & $+0.0009$\\
    \bottomrule
    \end{tabular}%
\end{table}

\begin{figure}[t]
    \centering
    \includegraphics[width=\linewidth]{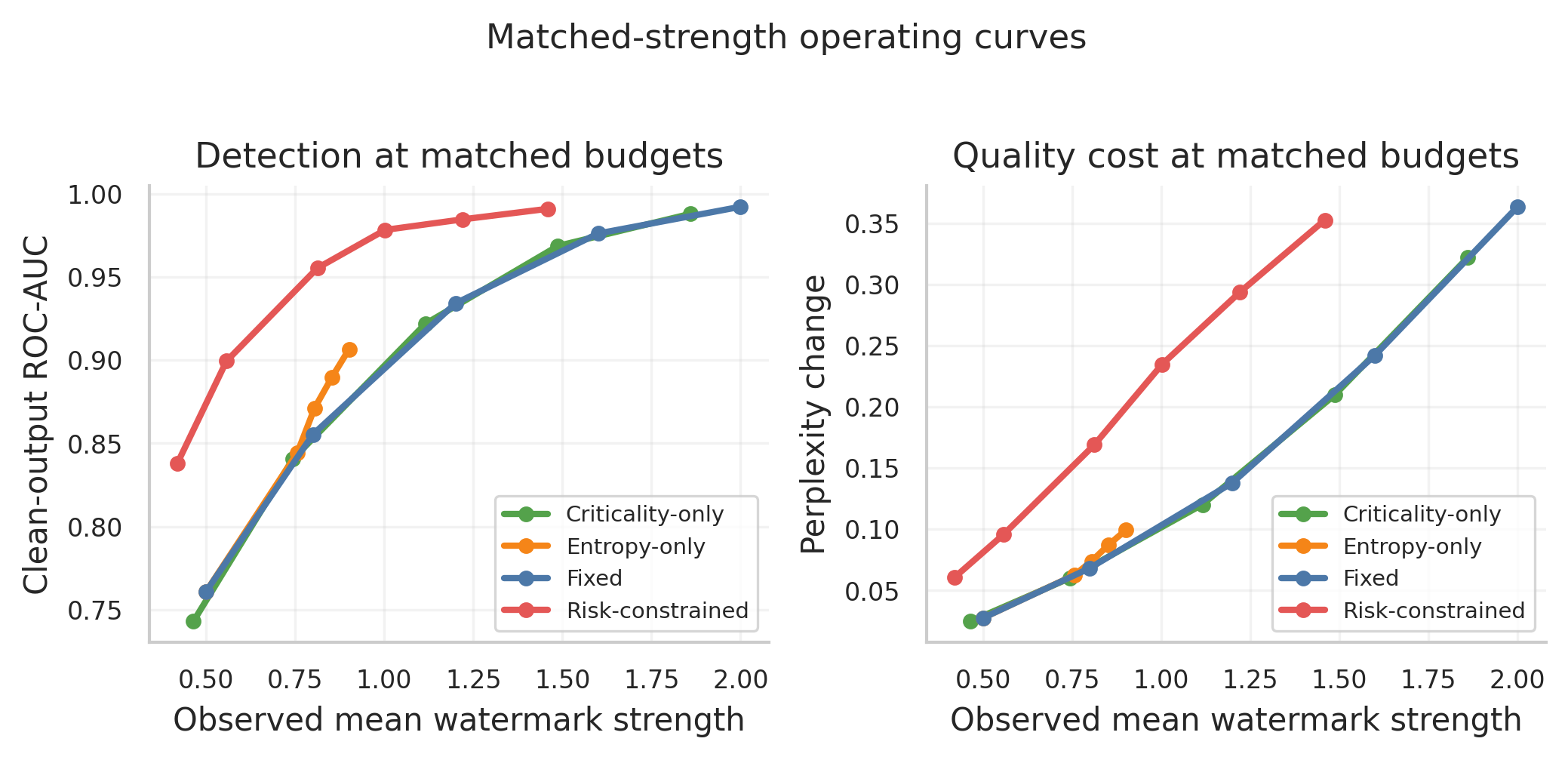}
    \caption{Matched-strength operating curves across watermarking strategies. The left panel shows clean-output ROC-AUC as a function of observed mean watermark strength. The right panel shows perplexity change as a function of observed mean watermark strength.}
    \label{fig:matched-budget-frontier}
\end{figure}

Figure~\ref{fig:matched-budget-frontier} visualizes the complete
matched-strength curves. The left panel shows clean-output ROC-AUC as a function
of observed mean watermark strength; the right panel shows perplexity change as
a function of observed mean watermark strength. The curves illustrate the
detection efficiency of risk-constrained allocation: at the same mean strength,
the risk-constrained policy lies above the other curves on detection and above
them on perplexity change, matching the mechanism described above.

\subsection{Risk-Penalty Operating Curve}
\label{sec:lambda-sweep}

The risk penalty $\lambda$ provides continuous control over the operating point.
As $\lambda$ increases, mean watermark strength decreases, which generally
improves response similarity and reduces perplexity change, but also reduces
detection performance. Table~\ref{tab:lambda-sweep} reports the complete
$\lambda$ sweep.

\begin{figure}[t]
    \centering
    \includegraphics[width=\linewidth]{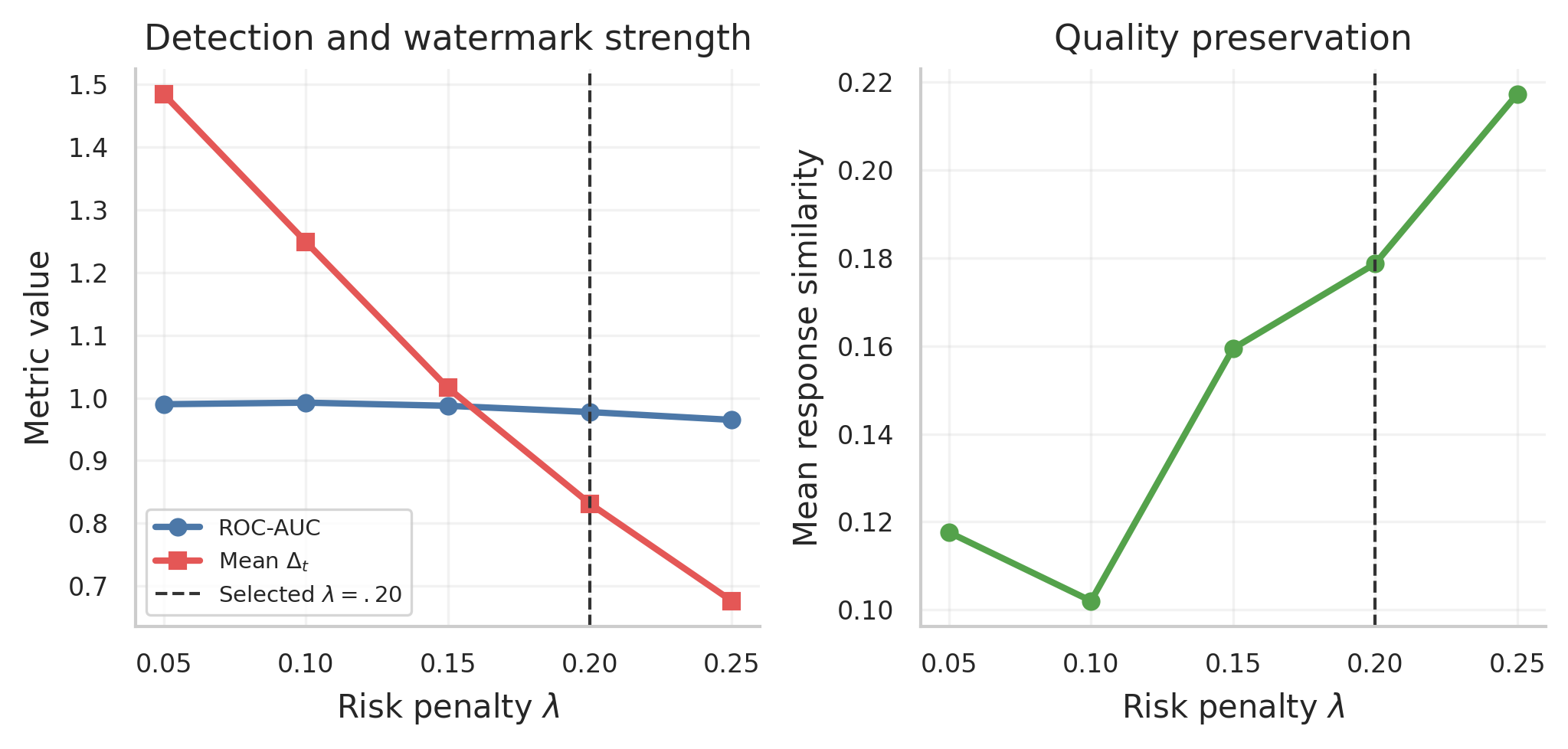}
    \caption{Calibration trade-off across risk penalties. The selected configuration uses $\lambda=0.20$ as an intermediate operating point between detection strength and response preservation.}
    \label{fig:lambda-calibration}
\end{figure}

For example, $\lambda=0.05$ produces mean strength $1.459$ and ROC-AUC
$0.9908$, whereas $\lambda=0.40$ produces mean strength $0.420$ and ROC-AUC
$0.8380$. The selected value $\lambda=0.20$ produces mean strength $0.812$ and
ROC-AUC $0.9552$, an intermediate operating point. The independent calibration
results are shown in Figure~\ref{fig:lambda-calibration}. The selected value is
therefore an intermediate operating point rather than an empirically optimal
value for every metric: it occupies the region of the curve where detection
remains high while intervention is substantially reduced, but a deployment that
prioritizes quality over detection could select a larger $\lambda$ and obtain
lower perplexity change at lower detection.

\begin{table}[t]
    \caption{Risk-constrained operating points across $\lambda$.}
    \label{tab:lambda-sweep}
    \centering
    \Large
    \setlength{\tabcolsep}{2.5pt}
    \begin{tabular}{@{}rrrrr@{}}
    \toprule
    $\lambda$ & Mean $\Delta$ & AUC & $\Delta$PPL & Similarity\\
    \midrule
    0.05 & 1.459 & 0.9908 & 0.3524 & 0.1177\\
    0.10 & 1.221 & 0.9845 & 0.2937 & 0.1367\\
    0.15 & 1.003 & 0.9782 & 0.2347 & 0.1736\\
    0.20 & 0.812 & 0.9552 & 0.1689 & 0.2087\\
    0.30 & 0.558 & 0.8994 & 0.0956 & 0.2706\\
    0.40 & 0.420 & 0.8380 & 0.0605 & 0.3195\\
    \bottomrule
    \end{tabular}%
\end{table}

\begin{figure}[t]
    \centering
    \includegraphics[width=\linewidth]{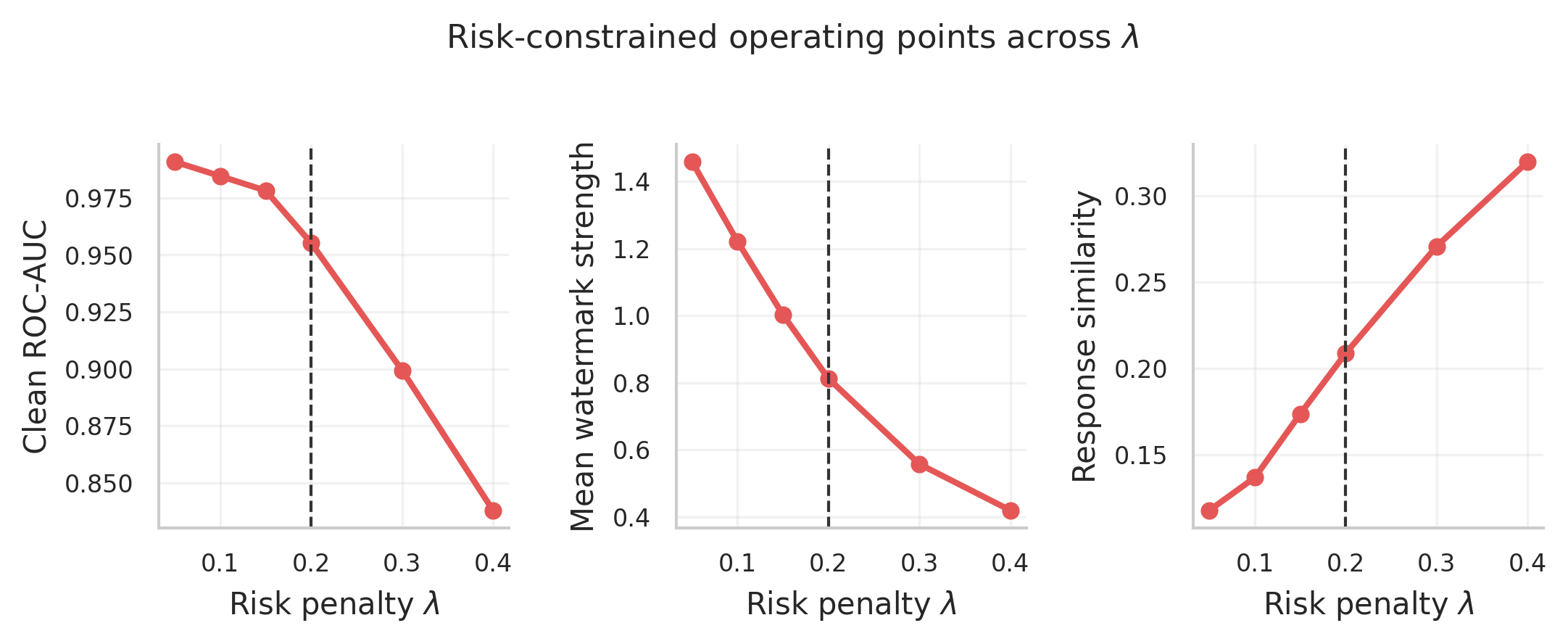}
    \caption{Risk-constrained operating points across $\lambda$. Increasing $\lambda$ reduces mean watermark strength and generally improves response similarity, while decreasing detection performance. The dashed line marks the selected value $\lambda=0.20$.}
    \label{fig:lambda-operating-curve}
\end{figure}

\subsection{Ablation Interpretation}
\label{sec:ablation}

The ablation separates three distinct behaviors. Criticality-only allocation
achieves near-fixed detection ($0.9879$ vs. $0.9921$) at slightly lower
intervention ($1.860$ vs. $2.000$), indicating that the relevance signal is
conservative: it attenuates strength only on a subset of decisions, and
therefore preserves detection. Entropy-only allocation achieves the strongest
automatic quality values but the weakest detection ($0.9065$), indicating that
uncertainty alone is an insufficient detection-benefit signal. The joint rule
combines these behaviors. In the default comparison it lies between the
extremes on both axes (Tables~\ref{tab:detection} and~\ref{tab:quality}). At
matched strength (within $0.07$ of the joint rule's mean strength $0.812$) it achieves higher detection than either component alone
($0.9552$ at mean strength $0.812$; Table~\ref{tab:matched-strength}). Whether
the specific relevance construction contributes beyond a strength-matched
random control remains an open question and is the natural next ablation; the
matched-strength result establishes that the joint rule is not merely the sum
of its parts in the default comparison, but isolating the relevance term
requires a randomized control matched for strength distribution.

\subsection{Paraphrase and Synonym Robustness}
\label{sec:robustness}

Table~\ref{tab:paraphrase} reports clean and paraphrased detection performance.
Under the evaluated paraphrase procedure, risk-constrained ROC-AUC decreases
from $0.9552$ to $0.9459$ and F1 decreases from $0.8929$ to $0.8769$. All
methods retain paraphrased ROC-AUC above $0.89$, with the smallest degradation
observed for fixed watermarking ($0.9921\to0.9885$) and the largest for
entropy-only ($0.9065\to0.8940$). A measurable watermark signal therefore
survives the tested transformation for all methods.

The synonym-substitution results show a monotone relationship between mean
watermark strength and detector-score degradation: the risk-constrained method,
with the lowest mean strength ($0.812$), exhibits a larger detector-score
reduction than fixed ($2.000$) or criticality-only ($1.860$). This is
consistent with the operating-characteristic view of Section~2: lower-intervention
operating points trade detection robustness for reduced perturbation, and the
paraphrased ROC-AUC remains above $0.94$ for the risk-constrained method.

\begin{table}[t]
    \caption{Clean and paraphrased detection performance on 1,000 instances.}
    \label{tab:paraphrase}
    \centering
    \Large
    \setlength{\tabcolsep}{2pt}
    \begin{tabular}{@{}lrrrrrr@{}}
    \toprule
    Method & Clean AUC & Para. AUC & Clean F1 & Para. F1 & Clean Recall & Para. Recall\\
    \midrule
    Fixed & 0.9921 & 0.9885 & 0.9370 & 0.9340 & 0.9890 & 0.9830\\
    Entropy-only & 0.9065 & 0.8940 & 0.8102 & 0.7864 & 0.7640 & 0.7270\\
    Criticality-only & 0.9879 & 0.9837 & 0.9340 & 0.9289 & 0.9830 & 0.9730\\
    Risk-constrained & 0.9552 & 0.9459 & 0.8929 & 0.8769 & 0.9050 & 0.8760\\
    \bottomrule
    \end{tabular}%
\end{table}

\begin{figure}[t]
    \centering
    \includegraphics[width=\linewidth]{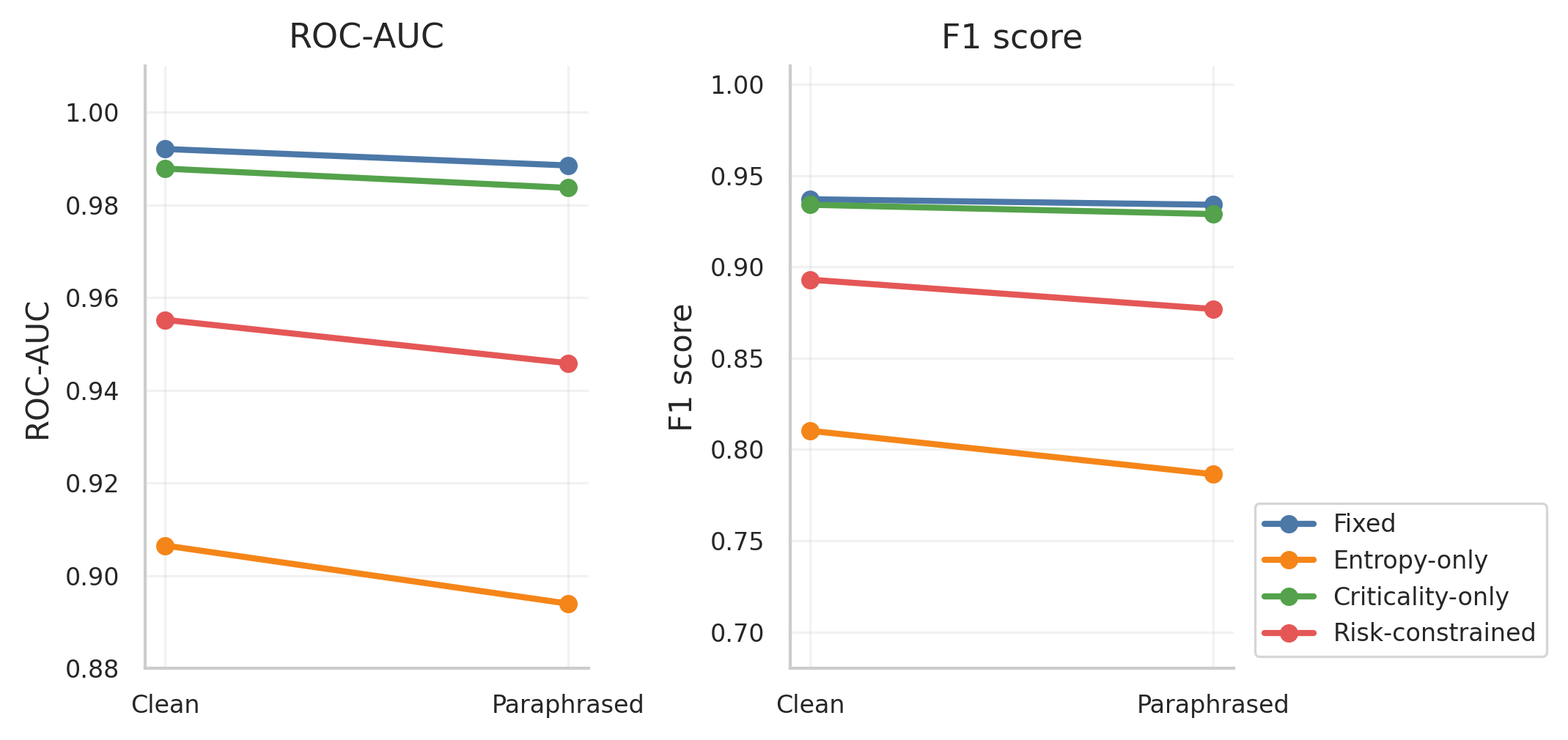}
    \caption{Detection stability under the evaluated paraphrase transformation. Lines connect clean and paraphrased ROC-AUC and F1 values for each watermarking strategy.}
    \label{fig:paraphrase-robustness}
\end{figure}

These results characterize robustness under the evaluated transformations only.
They do not establish resistance to arbitrary paraphrasing, deletion,
truncation, repeated rewriting, or adaptive detector-aware attacks.

\subsection{Detector-Score Distributions}
\label{sec:detector-ecdf}

\begin{figure}[t]
    \centering
    \includegraphics[width=\linewidth]{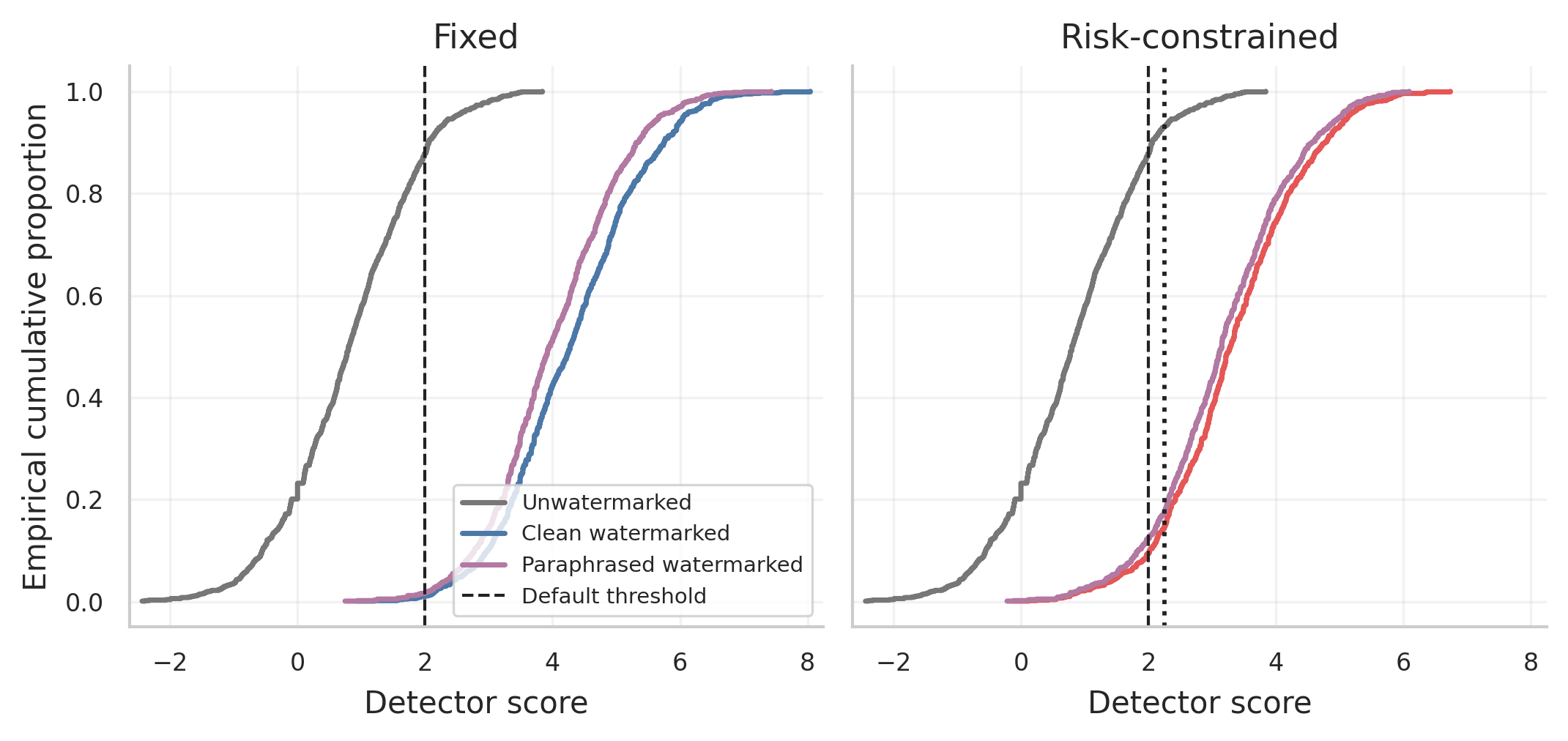}
    \caption{Empirical detector-score distributions for unwatermarked, clean-watermarked, and paraphrased-watermarked responses. The dashed line indicates the default detector threshold.}
    \label{fig:detector-ecdf}
\end{figure}

Figure~\ref{fig:detector-ecdf} shows empirical detector-score distributions for
unwatermarked, clean-watermarked, and paraphrased-watermarked responses. The
watermarked responses remain separated from the unwatermarked responses,
although paraphrasing shifts the distribution toward lower detector scores.
The dashed line indicates the default detector threshold; the shift under
paraphrasing is consistent with the ROC-AUC changes reported in
Table~\ref{tab:paraphrase} and with the threshold-calibration results in
Section~\ref{sec:threshold}.

\subsection{Threshold Calibration}
\label{sec:threshold}

Repeated held-out calibration shows that a nominal $5\%$ false-positive-rate
target produces empirical FPRs with mean $4.84\%$, standard deviation $1.82$
percentage points, and a range of approximately $2.38\%$--$8.62\%$ across
$R=20$ random $200/800$ splits. This is the most operationally
consequential result in this section: a single nominal threshold does not
guarantee the same empirical FPR on future traffic, and the width of the
distribution should be reported alongside detection performance.

The calibrated thresholds and their held-out performance are reported in
Table~\ref{tab:threshold}. For the nominal $5\%$ target, clean recall has mean
$77.79\%$ (standard deviation $5.23$ points) and paraphrased recall has mean
$73.56\%$ (standard deviation $6.42$ points). Clean and paraphrased F1 have
means of approximately $0.851$ and $0.823$, respectively. We therefore
recommend that provenance systems report calibration variance rather than a
single operating point, and recalibrate on representative traffic.

\begin{figure}[t]
    \centering
    \includegraphics[width=\linewidth]{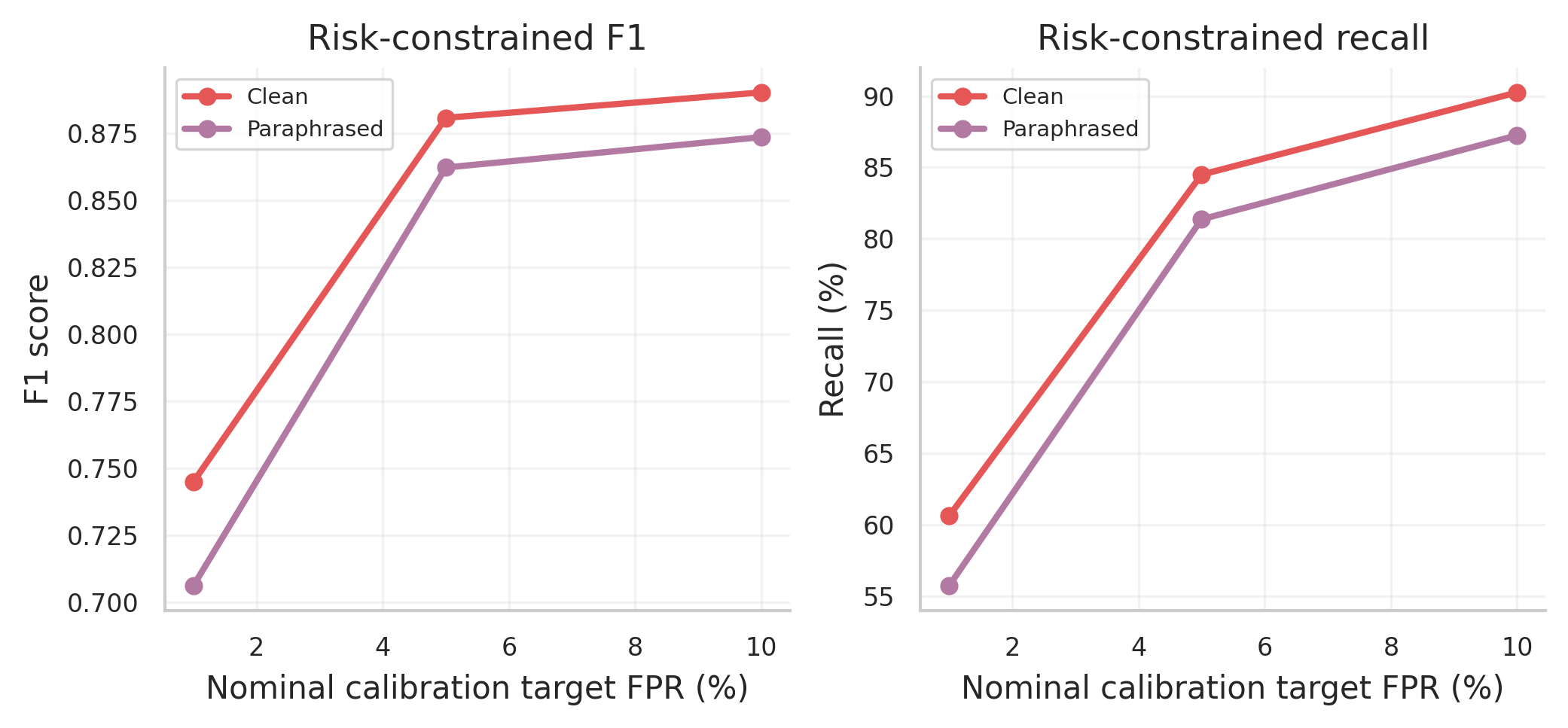}
    \caption{Detection operating points after held-out threshold calibration. The figure illustrates the relationship between the selected false-positive-rate target, empirical detection behavior, and recall.}
    \label{fig:threshold-operating-points}
\end{figure}

\begin{table}[t]
    \caption{Risk-constrained held-out threshold calibration.}
    \label{tab:threshold}
    \centering
    \Large
    \setlength{\tabcolsep}{2pt}
    \begin{tabular}{@{}rrrrrrr@{}}
    \toprule
    Target FPR & Threshold & Test FPR & Clean AUC & Clean F1 & Para. AUC & Para. F1\\
    \midrule
    0.01 & 3.0096 & 0.0213 & 0.9524 & 0.7450 & 0.9422 & 0.7063\\
    0.05 & 2.2549 & 0.0738 & 0.9524 & 0.8808 & 0.9422 & 0.8623\\
    0.10 & 2.0125 & 0.1250 & 0.9524 & 0.8903 & 0.9422 & 0.8736\\
    \bottomrule
    \end{tabular}%
\end{table}

Taken together, the results show that the proposed method exposes a
controllable security--quality frontier. Its main advantage is not universal
improvement on every metric, but the ability to select different operating
points by changing $\lambda$ and the resulting token-level watermark strength.

\section{Discussion}
\label{sec:discussion}

The experiments support a narrower but more useful conclusion than uniform
superiority over existing VLA watermarking methods: risk-aware allocation
exposes a controllable operating frontier, and different allocation policies
occupy distinguishable regions of that frontier.

\paragraph{Operating regimes.}
The results allow us to characterize which allocation policy is preferable
under different deployment requirements. A deployment that prioritizes
detection above all other considerations is best served by fixed or
criticality-only watermarking, which achieve ROC-AUC above $0.98$ at mean
strength above $1.8$. A deployment that prioritizes response quality and
tolerates weaker provenance evidence is best served by entropy-only
watermarking, which achieves the strongest automatic quality metrics at the
cost of ROC-AUC $0.9065$. A deployment that requires strong detection under a
constrained intervention budget—for example, one that must operate near a
maximum allowable perplexity increase or response-similarity floor—is best
served by risk-constrained allocation, which reaches ROC-AUC $0.9552$ at
matched mean strength where fixed, entropy-only, and criticality-only policies
reach $0.8553$, $0.8442$, and $0.8405$. This last regime is the one in which
the proposed method has a concrete operational advantage, and it is the regime
in which the matched-strength analysis matters most.

\paragraph{Controllability as the design contribution.}
The main design contribution is not a new watermark in the sense of a new
partition rule or a new detector, but a policy parameter $\lambda$ that
traverses a family of allocation behaviors while holding the model, the
partition mechanism, and the detector fixed. This matters for deployment
because it allows a practitioner to select an operating point on representative
traffic after training, without retraining or re-engineering the generation
pipeline. Fixed and entropy-only policies cannot expose this flexibility: fixed
watermarking has one operating point, and entropy-only watermarking exposes a
one-dimensional quality-detection curve that is not responsive to a deployment
constraint on mean intervention. The \(\lambda\)-controlled frontier is
therefore a practical capability, not a numerical improvement on a single
metric.

\paragraph{Scope of the relevance-risk signal.}
The ablation results in Section~\ref{sec:ablation} separate three behaviors:
criticality-only allocation preserves detection at slightly lower intervention,
entropy-only allocation achieves the strongest automatic quality but the
weakest detection, and the joint rule combines these behaviors and, at matched
strength, achieves higher detection than either component alone. This supports
the use of a joint signal but does not establish that the specific relevance
construction contributes beyond the strength distribution it induces. A
randomized-risk control matched for strength distribution is the natural next
ablation; until it is run, the relevance proxy should be treated as an
operational allocation signal rather than as a validated semantic-importance
estimator.

\paragraph{Image--text consistency.}
The CLIP results are essentially unchanged on average for all methods, with
the largest change being fixed watermarking at $-0.0042$ and entropy-only and
risk-constrained watermarking changing by less than $10^{-3}$. This should be
interpreted as evidence that the watermark does not substantially perturb the
selected automatic image--text metric, not as evidence of preserved visual
grounding. Human judgments, grounded question-answering accuracy,
object-relation accuracy, or downstream task performance would be required for
a stronger semantic-preservation claim, and these remain for future work.

\paragraph{Calibration as deployment practice.}
The calibration results in Section~\ref{sec:threshold} show that a nominal
false-positive-rate target is not a deployment guarantee: at a nominal $5\%$
target, empirical FPRs across repeated held-out splits have mean $4.84\%$ but
standard deviation $1.82$ percentage points and a range of approximately
$2.38\%$--$8.62\%$. This is a general property of finite-sample threshold
selection and not specific to the proposed method, but it has practical
consequences for provenance systems. A deployed detector should be calibrated
on representative unwatermarked traffic, its calibration variance should be
reported alongside its operating point, and re-calibration should be planned as
traffic distributions drift. Single-threshold deployment claims are unreliable
without this discipline.

The overall picture that emerges is that adaptive watermarking is useful
precisely because it lets a deployer choose which part of the security--quality
frontier to occupy. The contribution of the present work is to expose that
frontier in a controlled and reproducible way and to characterize where
different allocation policies sit on it.

\section{Limitations and Responsible-Use Considerations}
\label{sec:limitations}

\paragraph{Evaluation scope.}
The study evaluates one primary VLA configuration (LLaVA-1.5-7B), one
image--query dataset (a 1,000-instance AMBER subset), and one prompt template.
Results may depend on the model checkpoint, prompt format, decoding policy,
image domain, and detector implementation, and additional configurations are
needed to establish generality.

\paragraph{Proxy-metric limitation.}
Perplexity, response similarity, and CLIP are automatic proxy metrics. They
are informative about likelihood, textual agreement, and image--text alignment
under the chosen scorers, but they do not establish factual correctness, visual
grounding, user utility, or downstream task success. The study does not include
human judgments, grounded question-answering accuracy, or task-specific semantic
evaluation, and these are required before any strong semantic-preservation
claim.

\paragraph{Transformation scope.}
The robustness evaluation covers one synonym-substitution procedure and one
paraphrase procedure. In the current stage, it does not establish resistance to deletion,
truncation, repeated rewriting, adaptive paraphrasing, or detector-aware
attacks. Adaptive attacks optimized against the detector are a particularly
important threat that the current study does not evaluate.

\paragraph{Calibration variability.}
The threshold calibration uses finite samples (200 calibration instances and
800 held-out instances, repeated across splits). Reported FPRs should be
interpreted as estimates for the evaluated distribution, not as universal
deployment guarantees. Threshold selection should be recalibrated on
representative traffic before deployment.

\paragraph{Responsible use.}
A watermark is not proof of factual correctness, model safety, privacy, or
authorization by itself. A detected watermark should be treated as provenance
evidence with a known operating characteristic and used together with access
control, logging, content review, and other governance mechanisms. The system
should not be used to make high-impact decisions solely from a watermark score,
and a missed detection should not be treated as evidence that content was not
generated by the marked model, since detection is statistical and its
false-negative rate is nonzero at any useful operating point.

\section{Conclusion}
\label{sec:conclusion}

We studied risk-aware semantic watermark allocation for vision--language
agents. The method combines normalized predictive uncertainty with a zero-shot
multimodal relevance-risk proxy and derives token-level watermark strength from
a local detection-benefit versus risk objective. The risk penalty $\lambda$
exposes a continuous operating frontier that can be traversed at deployment
time without retraining the model or changing the detector.

Across 1,000 image--query instances, the default joint strategy reduced mean
watermark strength by approximately $59.4\%$ relative to fixed-strength
watermarking while reducing perplexity degradation and preserving a clean
ROC-AUC of $0.9552$ and a paraphrased ROC-AUC of $0.9459$. A matched-strength
analysis showed that this detection advantage is not merely a consequence of
lower average intervention: at comparable mean strength, risk-constrained
allocation achieves substantially higher ROC-AUC than fixed, entropy-only, and
criticality-only policies, at the cost of larger automatic quality
perturbation. This identifies risk-aware allocation as detection-efficient
rather than quality-dominant, and it shows that simpler policies cannot reach
the same operating points at equal watermark budget.

The central conclusion is therefore a controllable trade-off rather than
uniform dominance. Different allocation policies occupy distinguishable
regions of the security--quality frontier: fixed and criticality-only policies
provide stronger detection at high intervention, entropy-only allocation
provides stronger automatic quality preservation with weaker detection, and
the joint strategy offers an intermediate detection-efficient operating point
that is preferable when a deployment requires strong detection under a
constrained intervention budget. Calibration behavior reinforces the
deployment framing: finite-sample threshold selection produces substantial
false-positive-rate drift across splits, so provenance systems should report
calibration variance rather than a single operating point.

The results support risk-aware watermarking as a promising design direction
for trustworthy multimodal generation, while making the scope of the current
evidence explicit: the relevance-risk proxy is not yet validated as a
semantic-importance estimator, transformation robustness is established only
for the tested paraphrase and synonym-substitution procedures, and the
evaluation covers one VLA configuration and one dataset. Future work will
include a same-environment comparison with VLA-Mark, a randomized-risk control
matched for strength distribution, human and task-specific quality evaluation,
broader transformation and adaptive-attack evaluation, larger independent
calibration sets, and multiple VLA architectures and datasets.

\section*{GenAI Usage Disclosure}
The authors used a generative AI tool (ChatGPT Edu) solely for language polishing and structural editing of the manuscript. The AI toolwas  not used in the design of the study, the implementation of the watermarking method, the generation or analysis of experimental data, the production of figures or tables, or the interpretation of results. All technical content, experimental results, and claims are the authors' own. The authors have reviewed and edited all AI-assisted text and take full responsibility for the content of this paper.

\bibliographystyle{ACM-Reference-Format}
\bibliography{references}

\end{document}